# Framework elucidating the supersaturation dynamics of nanocrystal growth

Paul Z. Chen[†], Aaron J. Clasky[†], and Frank X. Gu[†,‡,*]

[†]Department of Chemical Engineering & Applied Chemistry, University of Toronto, Toronto, Ontario M5S 3E5, Canada
[‡]Institute of Biomedical Engineering, University of Toronto, Toronto, Ontario M5S 3G9, Canada

**ABSTRACT:** Supersaturation is the fundamental parameter driving crystal formation, yet its dynamics during the growth of colloidal nanocrystals (NCs) are poorly understood. Experimental characterization of supersaturation in colloidal syntheses has been difficult, limiting insight into the phenomena underlying NC growth. Hence, despite significant interest in the topic, how many types of NCs grow remain unclear. Here, we develop a framework to quantitatively characterize supersaturation *in situ* throughout NC growth. Using this approach, we investigate the seed-mediated synthesis of colloidal Au nanocubes, revealing a triphasic sequence for the supersaturation dynamics: rapid monomer consumption, sustained supersaturation, and then gradual monomer depletion. These NCs undergo different shape evolutions in different phases of the supersaturation dynamics. As shown with the Au nanocubes, we can use the supersaturation profile to theoretically predict the growth profile of NCs. We then apply these insights to rationally modulate shape evolutions, decreasing the yield of impurity NCs. Our findings demonstrate that the supersaturation dynamics of NC growth can be more complex than previously understood. While this study focuses experimentally on Au NCs, our framework is facile and applicable to a broad range of NCs undergoing classical growth. Thus, our methodology facilitates deeper understanding of the phenomena governing nanoscale crystal growth and provides insight towards the rational design of NCs.

## INTRODUCTION

Nanocrystals (NCs) are ordered materials with at least one dimension in the range of 1–100 nm and present physicochemical, electrical, magnetic, and optical properties that can be superior or unavailable to larger crystals.[1-3] For colloidal NCs, these properties depend on their size, shape, crystal habit, surface ligands, and elemental composition.[3-7] Using the mechanistic insight gained by elucidating reaction pathways, synthesis methods can grow colloidal NCs with precise control over these structural parameters,[8-10] potentiating the development of next-generation technologies in heterogenous catalysis, chemical processing, photonics, sensing, medicine, and environmental remediation.[11-18] Hence, a key focus in the fields of chemistry and nanoscience has been to understand how colloidal NCs grow.

Crystals can form classically or nonclassically. In classical crystal growth, monomeric units progressively integrate into crystal facets, as described by the terrace-step-kink (TSK) model of crystal surfaces.[19-21] In contrast, crystals grow nonclassically when monomers first form metastable intermediate clusters which then coalesce into a crystal.[22-24] While the distinction between classical and nonclassical growth is clear and presents a general scheme to classify crystal growth, the growth mechanisms of many colloidal NCs remain unclear. That is, how monomers integrate, or how intermediates attach and coalesce, into NCs is poorly understood, based in part on limited insight into the phenomena underlying NC growth.[25-26]

Supersaturation is the fundamental parameter that drives crystal formation and influences the growth mechanism, shape, exposed facets, crystal habit, size, and uniformity of NCs.[7, 27-30] It is defined as

$$\sigma = \ln\left(\frac{C}{C_\infty}\right) = \frac{\Delta\mu}{k_B T} \qquad (1)$$

where $C$ is the monomer concentration, $C_\infty$ is the saturation concentration of the monomer, $\Delta\mu$ is the chemical-potential difference between monomers and integrated units in the crystal, and $k_B T$ is the thermodynamic temperature. To form NCs, synthesis formulations modulate $C$ above the saturation (enabling growth if nuclei are present) or nucleation limits. For substrate-bound NCs, experiments can use open reactions and introduce vapor or liquid sources with fixed supersaturations to uncover mechanistic insight into growth.[30-33] Conversely, colloidal syntheses typically use closed conditions, and precursor is converted to monomer throughout the reaction (Figure 1a). These processes are dynamic, and previous approaches have required insight into the molecular structure of the monomer and analytical techniques, such as synchrotron-based time-resolved X-ray absorption near edge spectroscopy (XANES), to analyze supersaturation during growth.[34]

Thus, supersaturation dynamics have yet to be quantitatively characterized during the formation of many types of colloidal NCs, hindering mechanistic understanding. Since 1950, the LaMer model has guided a qualitative, but incomplete, understanding of how monomer tends to progress during the two-step homogenous nucleation and growth of colloidal NCs (Figure 1b).[35] The Finke-Watzky model[36] describes another two-step nucleation and growth process, in which slow, continuous nucleation occurs below the LaMer nucleation limit for monomer concentration ($C_{nu}$ in Figure 1b) before autocatalytic surface growth.[36] While Finke-Watzky nucleation deviates mechanistically from classical nucleation, these three-dimensional nuclei still form above a critical size described by classical theory.[36-37] Moreover, both models describe growth classically. That is, even when growth is kinetically controlled, NCs grow via the progressive integration of monomeric units, layer-by-layer growth relies on two-dimensional nucleation above a critical size on a facet, and NC facets have TSK structures.[20-21, 35-39] Unlike during nucleation, these models assume that the bulk, or local, supersaturation dynamics during growth generally follow a simple profile, diminishing steadily as the NCs grow (see stage III in Figure 1b). However, as revealed in this study, the supersaturation dynamics of colloidal NC growth can be more complex than described by these models.

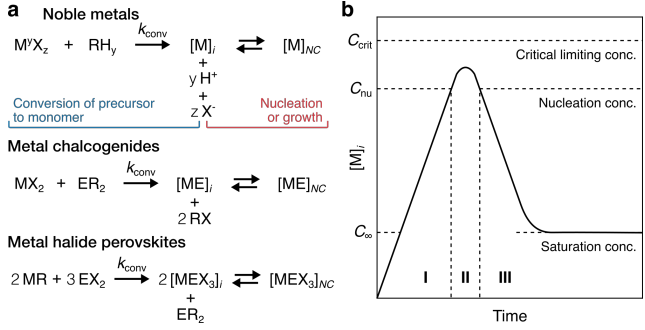

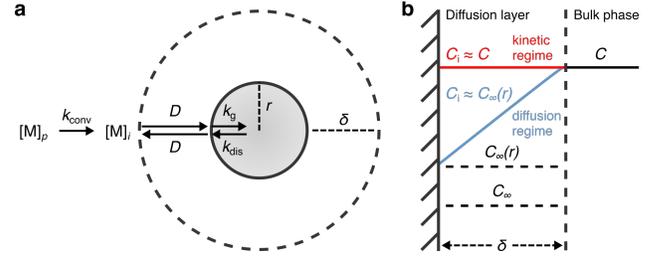

**Figure 1.** Precursor-limited homogeneous crystallization of colloidal NCs. (a) Schemes of the general reaction pathways to synthesize NCs composed of noble metals, metal chalcogenides, or metal halide perovskites. M = metal, E, X = counterions, R = reactant, y = oxidation state. The pathway for perovskites was represented by the oleate injection method for ternary NCs. (b) Qualitative LaMer model of the monomer concentration $[M]_i$ during NC crystallization, in which the monomer concentration reaches saturation and increases in supersaturation (I) before the two-step process for homogenous nucleation (II) and growth (III).

Here, we develop a framework to characterize supersaturation *in situ* throughout the growth of colloidal NCs. To avoid experimental measurements that disturb the reaction, this approach noninvasively analyzes the size-dependent optical properties of NCs to determine quantitative, temporal profiles of the supersaturation dynamics. These profiles facilitate theoretical modeling of, and mechanistic insight into, NC formation. By applying our framework to investigate the growth of Au nanocubes, we find that the supersaturation dynamics of NC growth can be more complex than previously understood via the LaMer or Finke-Watzky models. We then use the supersaturation profiles with theoretical equations to predict the growth profile of the nanocubes and identify supersaturation-associated NC shape evolutions. Based on these insights, we rationally modulate the shape evolutions of the nanocubes during growth.

## RESULTS AND DISCUSSION

**Framework to characterize supersaturation *in situ* during NC growth.** To develop the framework, we first consider the dynamics of monomer during colloidal NC growth, as summarized in Figure 2. Colloidal NCs undergoing classical growth typically involve precursor-limited reactions,[38-40] in which monomer is converted from precursor before it incorporates into a NC (Figure 1a). This predominantly occurs through one of two pathways: conversion occurs in the bulk phase and monomers diffuse through the stagnant layer of liquid surrounding the NC, or precursors diffuse through the stagnant layer so that the NC surface can catalyze the conversion to monomer.[35-37]

**Figure 2.** Schematic representations of monomer dynamics during colloidal NC growth. (a) The kinetic processes governing classical growth for a colloidal NC of radius $r$. Monomer or precursor diffuses from the bulk phase through the stagnant layer before reaching the NC, where monomer can incorporate into the crystal. $[M]_p$ represents the precursor, $[M]_i$ represents the monomer, $k_{conv}$ is the rate constant for conversion of precursor to monomer, $D$ is the bulk diffusion coefficient of the monomer, $k_g$ is the rate constant for growth, $k_{dis}$ is the rate constant for dissolution, and $\delta$ is the thickness of the stagnant layer surrounding the NC. (b) Monomer concentration relative to the NC surface (striped surface). The rate-controlling step for growth can be the integration of monomer into the crystal (kinetic regime) or the diffusion of monomer to the NC (diffusion regime). $C$ is the bulk concentration of precursor (for autocatalytic surface growth) or monomer, $C_i$ is the monomer concentration at the NC interface, $C_\infty$ is the bulk saturation concentration, and $C_\infty(r)$ is the local saturation concentration at NC interface, as influenced by the Gibbs-Thomson effect.

At the interface of the bulk phase with a NC, the local saturation concentration of monomer increases relative to that of a macroscopic crystal, which is described by the Gibbs-Thomson equation:[19]

$$C_\infty(r) = C_\infty \exp\left(\frac{2v_c\gamma}{rRT}\right) \approx C_\infty\left(1 + \frac{2v_c\gamma}{rRT}\right) \quad (2)$$

where $r$ is the radius of a spherical NC, $v_c$ is the molar volume of the monomer, $\gamma$ is the surface energy and $R$ is the gas constant. The monomer concentration at the NC interface $C_i$ is then[41]

$$\frac{C_i - C_\infty(r)}{C - C_i} = \frac{D}{rk_0^g}\left(1 + \frac{r}{\delta}\right) \quad (3)$$

where $D$ is the bulk diffusion coefficient of the monomer or precursor that traverses the stagnant layer, $k_0^g$ is the rate constant for growth of a flat ($r \to \infty$) interface, and $\delta$ is the thickness of the stagnant layer. For NCs, $\delta$ tends to be much larger than $r$, and $C_i$ can be considered in two regimes. In the diffusion regime ($D \ll rk_0^g$), the rate-controlling step for growth is the diffusion of monomer or precursor through the stagnant layer, and $C_i \approx C_\infty(r)$. Conversely, in the kinetic regime ($D \gg rk_g$), the rate-controlling step is the incorporation of monomer into the crystal, and $C_i \approx C$ (Figure 2b).

Since directly measuring the concentration of bulk monomer in colloidal NC syntheses is experimentally difficult, we sought to find another approach to characterize supersaturation. In classical growth, the rate at which monomers incorporate into crystals, $R_{NC}(t)$, is a function of supersaturation,[19] which has been demonstrated in the synthesis of colloidal NCs.[34, 38-39] Furthermore, colloidal NCs show a broad range of size-dependent optical properties, such as light scattering, absorbance, and fluorescence, which can be measured in real time by facile, noninvasive techniques.[42-44] Our framework will interpret the optical progression of the NCs throughout growth to determine $R_{NC}(t)$ and then $\sigma(t)$.

To obtain an expression for $R_{NC}(t)$, we start with an expression based on eqs (2) and (3). As obtained by Sugimoto[41] and expanded by Talapin et al.[45], the radial rate of classical growth for a colloidal NC is described by

$$\frac{dr}{dt} = v_c D C_\infty \left[ \frac{\exp(\sigma) - \exp\left(\frac{2v_c \gamma}{rRT}\right)}{r + \frac{D}{k_0^g} \exp\left(\alpha \frac{2v_c \gamma}{rRT}\right)} \right] \quad (4)$$

where $\alpha$ is the transfer coefficient of growth in the reaction. Fick's first law relates growth rate to the total flux of monomer onto the surface of a single NC:[41]

$$J = \frac{4\pi r^2}{v_c} \frac{dr}{dt} \quad (5)$$

Hence, the intensive form of the molar rate of monomer consumption can be formulated as

$$R_{NC}(t) = \frac{n_{NC}(t)}{V_s} J(t) \quad (6)$$

where $n_{NC}(t)$ is the number of NCs undergoing growth and $V_s$ is the volume of the synthesis formulation. During nucleation reactions, $n_{NC}(t)$ changes as nuclei emerge or disappear. On the other hand, seed-mediated syntheses spatiotemporally separate growth from nucleation, and these growth reactions may take $n_{NC}$ as the number of seeds added, a constant parameter.

In the diffusion regime, combining eqs (4)-(6) leads to an expression for supersaturation:

$$\sigma_d(t) = \ln \left[ \frac{V_s}{4\pi n_{NC}(t) D C_\infty r(t)} R_{NC}(t) + \exp\left(\frac{2v_c \gamma}{r(t)RT}\right) \right] \quad (7)$$

The term $2v_c\gamma/RT$ tends to be on the order of 1 nm, meaning $r > 10$ nm leads to a small contribution from the Gibbs-Thomson effect. In this case, NCs grow at rates comparable to bulk crystals and

$$\sigma_d(t) = \ln \left[ \frac{V_s}{4\pi n_{NC}(t) D C_\infty r(t)} R_{NC}(t) + 1 \right] \quad (8)$$

In the kinetic regime, supersaturation can be derived similarly as

$$\sigma_k(t) = \ln \left[ \frac{V_s}{4\pi n_{NC}(t) k_g C_\infty r^2(t)} R_{NC}(t) + 1 \right] \quad (9)$$

Eqs (7)-(9) express $\sigma(t)$ as a function of $R_{NC}(t)$. As mentioned previously, we can determine a second expression for $R_{NC}(t)$ from the size-dependent optical properties of NCs. For instance, the optical attenuation through a suspension of metal NCs can be measured during a reaction and is described by the Beer-Lambert Law:

$$A = cl\varepsilon \quad (10)$$

where $c$ represents either the concentration of particles $n_{NC}/V_s$ or the concentration of monomer in the NCs $C_{NC}$ depending on the reaction, $l$ is the sample path length, and $\varepsilon$ is the extinction coefficient, which is composed of the absorption ($\varepsilon_{abs} \propto r^3$) and scattering ($\varepsilon_{sc} \propto r^6$) coefficients.[46]

Synthesis can affect both $c$ and $\varepsilon$. When changes in $\varepsilon$ are small, such as during some crystallization reactions with minor changes in NC size,[47] $\varepsilon$ may be approximated as a constant value,[38-39] meaning $c = C_{NC}$ is the sole variable during the reaction and $A$ corresponds directly with $C_{NC}$. On another hand, seed-mediated syntheses involve a constant $c = n_{NC}/V_s$ but often grow NCs to larger sizes, meaning that $\varepsilon$ changes over the reaction. When absorption predominates the overall extinction, $\varepsilon$ is proportional to the volume of the NCs,[46] which depends on the amount of consumed monomer $C_{NC}$. Thus, $A$ corresponds with $C_{NC}$ in both cases.

When a synthesis reaction is run to completion, the final monomer concentration approaches $C_\infty$. If the total concentration of usable monomer added to the closed reaction is $C_0$, then the final concentration of consumed monomer is approximately $C_0 - C_\infty$. When colloidal NC syntheses directly introduce monomer or quantitatively convert precursor to monomer before seed addition, reactions typically follow pseudo-first-order kinetics.[40, 48-50] In both cases discussed above, $A$ corresponds with $C_{NC}$, and these pseudo-first-order kinetics can be described as

$$C_{NC}(t) = (C_0 - C_\infty) - (C_0 - C_\infty) \exp(-k_g t) \quad (11)$$

where $k_g$ is the rate constant of NC growth fitted on the profile of absorbance measurements taken throughout synthesis, and

$$R_{NC}(t) = -\frac{dC_{NC}(t)}{dt} = k(C_0 - C_\infty) \exp(-k_g t) \quad (12)$$

In this case, monomer is found either in the bulk phase, stagnant layer, or NC. Hence, $C(t) = C_0 - C_{NC}(t)$ in these pseudo-first-order reactions, and the bulk monomer concentration can be described by

$$C(t) = (C_0 - C_\infty) \exp(-k_g t) + C_\infty \quad (13)$$

Conversely, when seed-mediated reactions have an induction period, during which precursor is converted into monomer including via autocatalytic reaction, they often show sigmoidal kinetics. These kinetics can be represented by the Boltzmann sigmoid function,[51]

$$C_{NC}(t) = (C_0 - C_\infty) - \frac{C_0 - C_\infty}{1 + \exp[k_g(t - t_0)]} \quad (14)$$

where $t_0$ denotes the sigmoidal inflection point, and

$$R_{NC}(t) = -\frac{dC_{NC}(t)}{dt} = \frac{k_g(C_0 - C_\infty) \exp[k_g(t - t_0)]}{\{1 + \exp[k_g(t - t_0)]\}^2} \quad (15)$$

Thus, supersaturation can be characterized via eq (7)-(9), with $R_{NC}(t)$ specified by eq (12) for pseudo-first-order reactions, eq (15) for low-order reactions with an induction

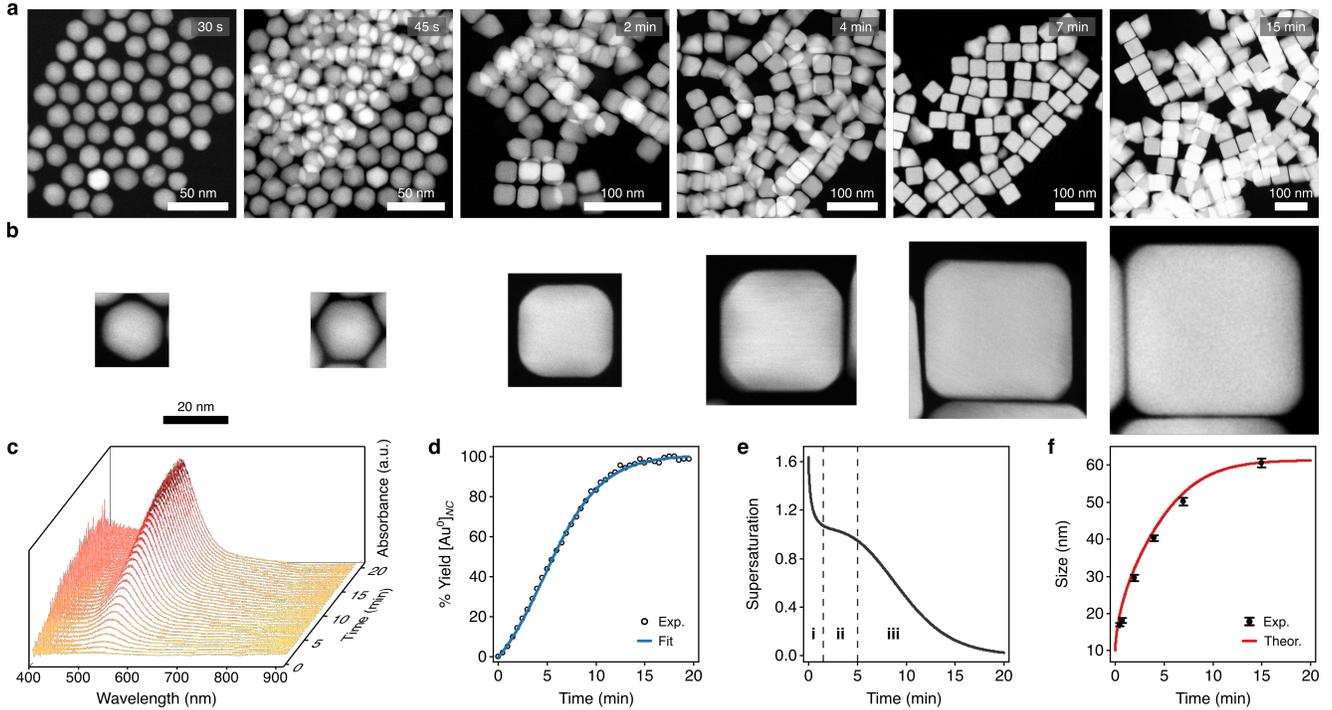

**Figure 3.** Characterization of the supersaturation-dependent growth of colloidal Au nanocubes. (a) HAADF-STEM images of the NCs found throughout colloidal nanocube growth. Times denote when growth was arrested relative to the initiation of growth. (b) HAADF-STEM images of single NCs from the above samples depicting the transition of cuboctahedra to cubes. The scale bar applies to each micrograph. (c) Time-resolved extinction spectra taken throughout colloidal NC growth. (d) Kinetics of NC growth analyzed via the peak nanoplasmonic extinction. The data were fitted to eq (14) ($r^2 > 0.99$). (e) Temporal profile of supersaturation dynamics throughout colloidal NC growth, as estimated by eqs (8) and (15). The dashed lines delineate the triphasic sequence of rapid monomer consumption (i), sustained supersaturation (ii), and then gradual depletion (iii). (f) Experimental and theoretically predicted growth of colloidal nanocubes. Theoretical growth was modelled using the supersaturation profile from (e) and eq (4). Dots and bars represent the mean and SD, respectively.

period, or higher-order equations for reactions with greater kinetic complexity[36, 40].

We can further formulate an expression to predict the final size of NCs undergoing growth. The concentration of crystallized monomer in a population of NCs can be approximated as $4\pi r^3 n_{NC}/(3v_c V_s)$, where $r$ is the volume-average radius. Since $C_{NC}(t \rightarrow \text{completion}) = C_0 - C_\infty$, the average NC size after seed-mediated growth can be estimated with

$$r_2^3 - r_1^3 = (C_0 - C_\infty)\frac{3v_c V_s}{4\pi n_{NC}} \quad (16)$$

where $r_2$ and $r_1$ are final and seed radii, respectively.

In summary, our framework analyzes the progression of a size-dependent optical property to quantify the bulk supersaturation dynamics throughout the classical growth of NCs, particularly those with simple geometries in its current form. In the interpretation above, the framework measures absorbance as this optical property and considers reactions in which absorption predominates the overall extinction. The latter condition is usually valid when colloidal NCs remain smaller than 50-60 nm.[42, 52] In-line spectrophotometers can rapidly noninvasively measure the absorbance through a transparent vial during synthesis, and high-frequency assessments of absorbance yield high-resolution supersaturation profiles. Moreover, since we characterize $\sigma(t)$ based on $R_{NC}(t)$, our approach requires neither insight into the source of the monomer nor its molecular structure, limiting the requirements of *a priori* knowledge and expanding the utility of the methodology. Based on the time-resolved profile of supersaturation in a growth reaction, we may theoretically predict the growth profile of the NCs and identify supersaturation-associated shape evolutions during synthesis, as we demonstrate below.

**Elucidating supersaturation dynamics throughout the growth of colloidal Au nanocubes.** We used our framework to investigate the growth of colloidal Au nanocubes. To assess how supersaturation dynamics influence the shape evolutions of NCs, we chose a seed-mediated synthesis formulation with cetyltrimethylammonium chloride (CTAC) that predominantly grows nanocubes but also leads to a small proportion of impurity shapes. To experimentally analyze NCs throughout the aqueous reaction, we arrested growth at various times using ligand exchange with thiol-terminated polystyrene and solvent transfer via tetrahydrofuran (THF) into toluene.[53] We imaged these samples using high-angle annular darkfield scanning transmission electron microscopy (HAADF-STEM) (Figure 3a), which showed that the cuboctahedral seeds primarily evolved to truncated cubes and then to cubes along growth (Figure 3b).

Towards a high-resolution assessment of the supersaturation dynamics, we took high-frequency absorbance measurements throughout the reaction (Figure 3c). The absorbance kinetics had an induction period, and we fitted them to eq (14) (the equation constants are summarized in Table S1; see section S1 in the Supporting Information for additional details). As shown in Figure 3c, this model showed a strong agreement with the experimental data (coefficient of determination, $r^2 > 0.99$). Moreover, since the absorbance of the NCs leads to visually discernable colors, we also took images of the synthesis formulation throughout growth and analyzed the colorimetric development (Figure S1a,b). Imaging and spectrophotometry yielded similar results in their analyses of the absorbance kinetics (Figure S1c), indicating that both approaches can be used to determine the variable parameters of eq (14).

We applied these fitted parameters to characterize the supersaturation dynamics in the reaction. When $D \ll 1$, colloidal NC growth tends to occur in the diffusion regime.[37] For our synthesis formulation, $D$ is on the order of $10^{-10}$ m$^2$ s$^{-1}$ (ref. [54]), and we used eqs (8) and (15) to determine the supersaturation dynamics (Figure 3e, the equation constants are summarized in Table S1). This revealed a triphasic sequence: an initial phase of rapid monomer consumption (0.0–1.5 min) followed by phases of sustained supersaturation (1.5–5.0 min) and then gradual monomer depletion (>5.0 min). Although seed-mediated syntheses of colloidal NCs have been studied heavily, the supersaturation dynamics of these reactions have been assumed to follow the descriptions of the growth stages of the LaMer (see section III in Figure 1b) or Finke-Watsky models, in which supersaturation diminishes steadily as the NCs grow.[35-36] Notably, the profile characterized throughout the growth of these colloidal nanocubes showed a triphasic sequence with distinct phases of monomer consumption (Figure 3e and Figure S2), indicating that the supersaturation dynamics of colloidal NC growth can be more complex than previously understood.

We next used the supersaturation profile (Figure 3e) to model the growth of the colloidal nanocubes using eq (4) (see section S2 in the Supporting Information for additional details). We also used the HAADF-STEM images of the NCs throughout the reaction to analyze the experimental sizes of the cuboctahedra and cubes (Figure S3 summarizes the sizing measurements). As shown in Figure 3f, the theoretical model showed a strong agreement with the experimental results, predicting the NC growth profile throughout the reaction. According to this profile, the NCs grew from 10.0 nm (the seed size[55-56]) to 29.0 nm in the initial phase of the supersaturation dynamics (0.0–1.5 min), then to 46.5 nm in the intermediate phase (1.5–5.0 min), and finally to approximately 61.2 nm (5.0 min to completion).

To further investigate the influence of the supersaturation dynamics on growth, we artificially simulated a supersaturation profile (Figure S4a) which lacks the initial phase of rapid monomer consumption and more resembles that expected when considering the LaMer model (see section S3 in the Supporting Information for additional details). We inputted the simulated supersaturation into eq (4) for another theoretical growth model, which predicted the final size of the NCs but did not conform to the growth profile during the earlier stages of growth (Figure S4b). This distinction reinforced that this reaction involves the complex

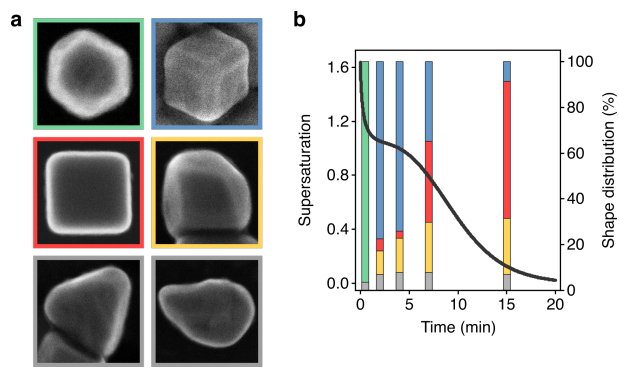

**Figure 4.** Supersaturation-associated shape evolutions of nanocubes. (a) SE-STEM images showing the NC shapes found throughout colloidal nanocube growth. Cuboctahedra (green), truncated cubes (blue), cubes (red), overgrown cubes (yellow), tetrahedra (grey) and other shapes (grey). The micrographs are not to scale. (b) Temporal profile of supersaturation dynamics taken from Figure 3e (black curve) overlaid on the distribution of shapes (bars) at 0.5, 2, 4, 7 and 15 min during colloidal growth. Shapes are denoted by the outline colors in (a).

supersaturation dynamics found in Figure 3e and further supported the utility of our framework.

**Supersaturation-associated shape evolutions of nanocubes during growth.** With the supersaturation profile in hand, we sought additional insight into the phenomena that facilitated growth by investigating the relationship between the supersaturation dynamics and NC shape evolutions. The HAADF-STEM images and additional scanning electron (SE)-STEM images identified six NC shapes present in this reaction (Figure 4a). Three of these shapes (cuboctahedra, truncated cubes, and cubes) were involved in the primary growth pathway in this reaction (Figure 3b), but the other three shapes (overgrown cubes, tetrahedra, and a small proportion of other, undefined shapes) were undesired. While cuboctahedral seeds grow into tetrahedra and the observed undefined shapes as kinetic impurities of synthesis,[55, 57-58] it was unclear how cube overgrowth occurred.

To study the relationship between the supersaturation dynamics and overgrowth, we used the SE-STEM images to analyze the distribution of NC shapes throughout growth (Table S2) and overlaid them on the supersaturation profile (Figure 4b). This showed that the population of cuboctahedra predominantly evolved to cubic shapes (truncated cubes, cubes, or overgrown cubes) during the initial phase of rapid monomer consumption. Truncated cubes were the dominant shape shortly after this transition but decreased in proportion as the reaction progressed. The overall yield of NCs with cubic shapes remained consistent throughout growth (92.9, 92.3, 92.1, and 93.0% at 2.0, 4.0, 7.0, and 15 min, respectively) (Table S2), suggesting that cubes and overgrown cubes emerged from the truncated cubes rather than directly from cuboctahedra.

Figure 4b also showed that cubes and overgrown cubes emerged in different phases of monomer consumption. Cubes primarily emerged in the third phase, gradual monomer depletion. Conversely, overgrowth primarily occurred in the first and second phases. By 2 min, 10.2% of the NCs were overgrown cubes. By 4 and 7 min, this proportion increased to 15.0 and 21.9%, respectively. Between 7-15 min, however, it only increased to 24.4%. That is, cube overgrowth was largely associated with the high supersaturation of the earlier phases of growth (Figure 4b). As described by Fick's First Law,[41, 45]

$$J = 4\pi Dr(C - C_i) \qquad (17)$$

higher supersaturation leads to higher rates of monomer flux to the NCs and, as our reaction occurs in the diffusion regime, higher rates of monomer incorporation. Excessive monomer incorporation into the NC surface is a plausible kinetic mechanism for overgrowth. Thus, we rationalized that decreasing supersaturation in the earlier stages of the reaction would decrease the yield of overgrown cubes.

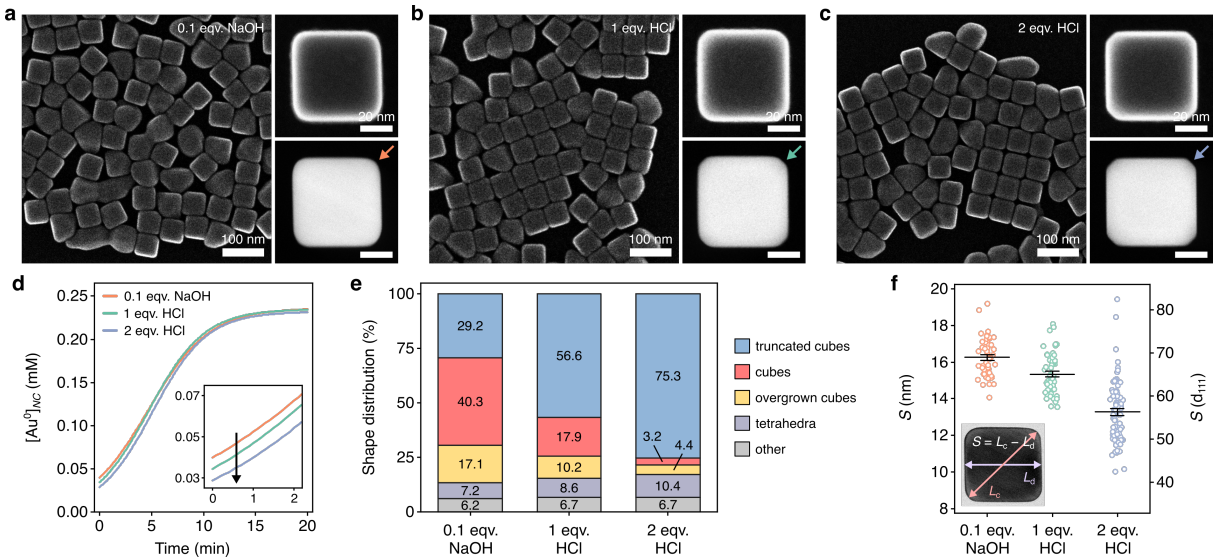

**Figure 5.** Supersaturation dynamics mediate shape evolutions during NC growth. (a-c) SE-STEM image of the NCs (left) and SE-STEM (top right) and HAADF-STEM (bottom right) image of an individual nanocube when grown with (a) 0.1 equivalents (eqv.) NaOH, (b) 1 eqv. HCl or (c) 2 eqv. HCl. The arrows point to nanocube corners, which increasingly truncate with greater amounts of HCl during growth. (d) Conversion kinetics for colloidal nanocube growth. Inset, early kinetics. (e) Shape yields for the various synthesis formulations. (f) Comparison of the corner sharpness of nanocubes from the various formulations. Inset, schematic representation of the measurements and sharpness index ($S$). $S$ is depicted in units of nm and {111} interplanar distances. Lines and whiskers represent means and 95% CIs, respectively.

**Modulating supersaturation dynamics to influence nanocube shape evolutions.** In our synthesis, the reduction of Au precursor to monomer produces $H^+$ and $Cl^-$ (Figure S5). The addition of HCl to this reaction should shift the reaction equilibrium and decrease the concentration of bulk monomer. Moreover, $H^+$ and $Cl^-$ are labile ions on Au surfaces and act minimally as facet-modifying agents.[25, 59] We hypothesized that adding HCl before the addition of seed would diminish monomer conversion in the early stages of growth and limit cube overgrowth. The synthesis formulation for the NCs shown in Figures 3 and 4 used 0.1 molar equivalents (eqv., relative to Au added) of NaOH. We compared these NCs to the NCs from two syntheses that replaced NaOH with 1.0 or 2.0 eqv. of HCl.

Figure 5a-c shows the NCs produced by these syntheses, with larger-area SE-STEM images shown in Figure S6. As done previously, we took high-frequency absorbance measurements throughout these reactions and analyzed the absorbance kinetics (Figure S7). Fitting them to eq (14) characterized the monomer conversion kinetics, which showed that the addition of HCl indeed diminished monomer conversion in the early stages of growth (Figure 5d). We then sized the NCs as depicted in Figure S1. As hypothesized, the additions of HCl diminished monomer conversion by shifting the reaction but still produced nanocubes of similar size (Table S3).

To investigate the influence of these different supersaturation dynamics, we analyzed the shape distribution for each sample. As shown in Figure 5e, the addition of HCl indeed coincided with lower yields of overgrown cubes. With 0.1 eqv. of NaOH added, 17.1% of the NCs were overgrown nanocubes. This proportion diminished to 10.2% for 1 eqv. of HCl and then to 4.4% for 2 eqv. of HCl. Moreover, the addition of HCl also coincided with increasingly truncated cubes (Figure 5a-c, e). Previous research has indicated that decreasing supersaturation during synthesis is expected to dull NC corners.[60] To further quantify this in our samples, we measured a sharpness index ($S$) for individual nanocubes, which is defined in the inset of Figure 5f. With 0.1 eqv. of NaOH added, the mean estimate of $S$ was 16.3 (95% confidence interval [CI]: 15.9-16.6) nm, whereas it was 15.3 (15.0-15.6) nm for 1 eqv. of HCl and 13.3 (12.9-13.7) nm for 2 eqv. of HCl. Taken together, these results show that altering the supersaturation dynamics during growth modulated the features and shape evolutions of these colloidal NCs.

## CONCLUSIONS

In conclusion, we developed a framework and the accompanying methodology to characterize supersaturation *in situ* throughout the classical growth of colloidal NCs. The approach can produce high-resolution profiles that describe the supersaturation dynamics of these reactions. Our study found that the growth of CTAC-capped Au nanocubes involves a triphasic sequence of monomer consumption, revealing that the supersaturation dynamics in NC syntheses can be more complex than previously understood, such as through the LaMer or Finke-Watsky models.[35-36] We further showed that the supersaturation profile can be used to theoretically predict the growth profile of NCs as well as identify supersaturation-associated shape evolutions. The emergence of cubes and overgrown cubes were associated with different phases and rates of monomer consumption. This provided insight towards rationally modulating the shape transitions of these NCs during growth. Moreover, since NCs generally interact with light through size-dependent properties,[38-39, 42-44] our facile approach can be employed directly or built upon, including better interpreting anisotropic growth, to broadly study how supersaturation progresses during and influences colloidal NC growth. Taken together, this work may facilitate the rational design of NCs and can be used to uncover insight into the complex thermodynamic and kinetic phenomena that mediate the growth of many types of NCs.

## EXPERIMENTAL SECTION

**Chemicals and Materials**. Gold(III) chloride hydrate (HAuCl4•xH2O, x ≈ 3; >99.995%), sodium bromide (NaBr, >99.99%), cetyltrimethylammonium bromide (CTAB, >99%), cetyltrimethylammonium chloride (CTAC, >99%), L-ascorbic acid (>99.0%), sodium borohydride (NaBH4, >99.99%), sodium hydroxide (NaOH, >98%), hydrochloric acid (HCl, 37%), thiol-terminated polystyrene (PS-thiol, 5 kDa, polydispersion index ≤1.1), tetrahydrofuran (THF, 99.9%) and toluene (99.8%) were purchased from Sigma-Aldrich (Oakville, ON, Canada). Unless otherwise specified, MilliQ water (18.2 MΩ.cm at 25 °C; Milli-Q® Reference Water Purification System; MilliporeSigma, Oakville, ON, Canada) was used for the experiments. Quartz cuvettes (20 ml) were purchased from Thermo Fisher Scientific (Ottawa, ON, Canada). Scintillation vials (20 ml) were purchased from VWR International (Mississauga, ON, Canada). Reagents were used as received.

**Nanoseed preparation.** The synthesis procedures were adapted from Park et al.[55] In a 20 ml scintillation vial, 10 mM HAuCl$_4$ (250 µl) was added to a solution of 100 mM CTAB (9.75 ml), followed by the rapid addition of ice-cold 10 mM NaBH$_4$ (600 µl). Samples were stirred at room temperature for 2 min at 1400 rpm and then placed in an incubator at 27°C for 3 h to nucleate nanoseeds. Samples were brown in color, indicating small colloidal NCs. To further grow the nanoseeds, 200 mM CTAC (2 ml), 100 mM ascorbic acid (1.5 ml), the above nanoseed solution (50 µl) and 0.5 mM HAuCl$_4$ (2 ml) were added sequentially to a 20 ml scintillation vial and stirred at room temperature for 15 min at 300 rpm. The color progressed to red and had an extinction peak around 520 nm, indicating growth to a diameter of 10 nm.[55-56] Samples were then twice centrifuged at 20,600xg for 30 min. The first resuspension was with MilliQ (1 ml) and the second was with 20 mM CTAC (1 ml).

**Nanocube synthesis.** In a 20 ml quartz cuvette, 100 mM CTAC (6 ml), 40 mM NaBr (30 µl), nanoseeds (30 µl), HCl or NaOH (200 µl of the solutions described below), 10 mM ascorbic acid (390 µl, dropwise) and 0.5 mM HAuCl$_4$ (6 ml) were added sequentially and then mixed via pipette. Aside from this initial pipette mixing step, the solution was left untouched for the rest of synthesis. The concentration of HCl or NaOH added was modified according to the experiment: 0.1 eqv. NaOH (1.5 mM), 1 eqv. HCl (15 mM) or 2 eqv. HCl (30 mM). Molar equivalents refer to the molar ratio between the added reagent and added Au.

**Arresting colloidal NC growth.** At various times during synthesis, the growth of colloidal NCs was arrested via ligand exchange and solvent transfer based on a procedure adapted from Park et al.[53] Nanocubes were prepared as previously described with 0.1 eqv. NaOH. At various time points, 8 ml of colloidal NCs was quickly added to a THF solution of 1.1 mM PS-thiol (10 ml) in a 20 ml scintillation vial. This solution was vortexed aggressively (>2 min) until it became grey and translucent. This procedure was used for each time point (30 s, 45 s, 2 min, 4 min, 7 min and 15 min). Samples were left at room temperature overnight to allow the colloidal NCs to settle in the vial. After this, most of the liquid was removed from the vial via pipetting, and vials were placed under reduced pressure until the liquid had evaporated (>2 h). Next, 1 ml toluene was added to each sample, followed by sonication to resuspend the NCs. Samples were then twice centrifuged at 21,000xg for 25 min. The first and second resuspensions were with 1.0 and 0.1 ml of toluene, respectively.

**Optical characterization of NC growth.** Nanocubes were synthesized in 20 ml quartz cuvettes (path length, 1 cm) as described above. To characterize growth via extinction, the cuvettes were placed in darkness in an apparatus, which passed light from a source (OceanOptics, Orlando, USA) using fiber optic cables (OceanView Optics, Orlando, USA), 10-m round subminiature assembly (SMA) connectors and detector collection lenses (L4 chamfered; OceanView Optics, Orlando, USA) to an in-line UV/Vis spectrometer (Flame T spectrometer; OceanView Optics, Orlando, USA) with a 4000-series detector (350-1000 nm filter; OceanView Optics, Orlando, USA). Extinction spectra (400-900 nm, step size: 0.217 nm) were collected every 30 s. The initial scan was taken immediately after HAuCl$_4$ was added which initiated growth, and this image was considered the start time (0 s) in kinetic analyses. To characterize growth via nanoplasmonic color, the cuvettes were imaged (EOS Rebel T7i with an EF 100 mm macro lens, Canon Canada, Inc., Toronto, ON, Canada) under fume hood lighting every 30 s or 1 min. The initial image was taken immediately after HAuCl$_4$ was added which initiated growth, and this image was considered the start time (0 s) in kinetic analyses.

**Scanning transmission electron microscopy.** STEM samples were prepared by drop casting the respective solution (5 µl) on a 400-mesh pure C, Cu grid (Ted Pella, Inc., Redding, USA) and dried under hood evaporation. Before drop casting, colloidal NCs were prepared via centrifugation and resuspension in MilliQ. Grids were cleaned with ultraviolet light (5-15 min per side) using a ZoneTEM (Hitachi High-Technologies Canada Inc, Etobicoke, ON, Canada) sample cleaner before imaging. HAADF-STEM and SE-STEM images were acquired in vacuo using a Hitachi HF-3300 300 kV Environmental TEM with an electron acceleration voltage of 300 kV.

**NC sizing and counting.** For sizing, two representative large-area HAADF-STEM images were analyzed using ImageJ (version 1.51s; National Institutes of Health, USA), yielding $n \geq 40$ NCs for each time point. Each NC that could be clearly delineated in its HAADF-STEM image was sized. For cubes, four measurements were taken for each NC: two corner-to-corner (C) and two edge-to-edge (E) measurements (Figure S3a). For cuboctahedra, three side-to-side (D) measurements were taken (Figure S3b). The data were presented as mean for each time point. To characterize the shape yields at each time point, each NC was assessed for its shape in four representative large-area HAADF-STEM images. For a perfectly sharp nanocube, $C = E\sqrt{2}$ based on the parameters described in Figure S3, where $C = \mathrm{smax}(C1, C2)$ and $E = \min(E1, E2)$. To delineate truncated cubes from cubes,[55] we applied a factor threshold of 1.3; that is, truncated cubes ($C < 1.3E$) were below this threshold, whereas cubes ($C \geq 1.3E$) were equal to or above it. The count for each shape was summed from these images, and yields were presented as percentages of the overall number of NCs counted. The sharpness index ($S$) of the nanocubes was measured in units of nm and $d_{111}$, where $d_{111}$ denotes {111} interplanar distances (the corners of face-centered cubic nanocubes point in the <111> direction[61]).

**Kinetics characterization and modeling.** The peak absorbance values between 500-600 nm were used to analyze the absorbance kinetics. The parameters used to characterize the kinetics of colloidal NC growth and to theoretically model growth are summarized in Table S2. Coefficients of determination ($r^2$) were calculated when fitting to $C_{\text{NC}}(t)$ from eq (14) for these reactions. Further information on the methods for kinetics characterization and modeling is described in the Supporting Information. Modeling was performed using Matlab R2021a (MathWorks, Inc, Natick, MA, USA).

## ASSOCIATED CONTENT

### Supporting Information

The Supporting Information is available free of charge on the ACS Publications website.

> Additional details for the analysis of growth kinetics, theoretically modeling of NC growth, and the parameters used are provided. Characterization of NC growth kinetics via nanoplasmonic color compared with extinction, schematic representations of NC sizing, simulated supersaturation dynamics without an initial period of and the resulting growth profile, and TEM images of nanocubes are also included (PDF).

## AUTHOR INFORMATION

### Corresponding Author

**Frank X. Gu** – *Department of Chemical Engineering & Applied Chemistry and Institute of Biomedical Engineering, University of Toronto, Toronto, Ontario M5S 3E5, Canada;* orcid.org/0000-0001-8749-9075; Email: f.gu@utoronto.ca

### Authors

**Paul Z. Chen** – *Department of Chemical Engineering & Applied Chemistry, University of Toronto, Toronto, Ontario M5S 3E5, Canada;* orcid.org/0000-0001-5261-1610

**Aaron J. Clasky** – *Department of Chemical Engineering & Applied Chemistry, University of Toronto, Toronto, Ontario M5S 3E5, Canada;* orcid.org/0000-0001-5455-1340

### Author Contributions

The manuscript was written through contributions of all authors.

### Notes

The authors declare no competing financial interests.

## ACKNOWLEDGMENTS

We thank J. Watchorn (UToronto) for helpful comments regarding this manuscript. STEM images were taken at the Ontario Center for the Characterization of Advanced Materials (OCCAM) at the University of Toronto. This work was funded by the Natural Sciences and Engineering Research Council of Canada (NSERC) Discovery grant (06441). P.Z.C. was supported by the NSERC Vanier Canada Graduate Scholarship. A.J.C. was supported by the Ontario Graduate Scholarship. F.X.G. was supported by the NSERC Senior Industrial Research Chair program.

## REFERENCES

1. Alivisatos, A. P., Semiconductor Clusters, Nanocrystals, and Quantum Dots. *Science* **1996,** *271* (5251), 933-937.
2. Shamsi, J.; Urban, A. S.; Imran, M.; De Trizio, L.; Manna, L., Metal Halide Perovskite Nanocrystals: Synthesis, Post-Synthesis Modifications, and Their Optical Properties. *Chem Rev* **2019,** *119* (5), 3296-3348.
3. Liu, H.; Di Valentin, C., Shaping Magnetite Nanoparticles from First Principles. *Phys. Rev. Lett.* **2019,** *123* (18), 186101.
4. Zhao, M.; Chen, Z.; Shi, Y.; Hood, Z. D.; Lyu, Z.; Xie, M.; Chi, M.; Xia, Y., Kinetically Controlled Synthesis of Rhodium Nanocrystals with Different Shapes and a Comparison Study of Their Thermal and Catalytic Properties. *J. Am. Chem. Soc.* **2021,** *143* (16), 6293-6302.
5. Pan, J.; Shang, Y.; Yin, J.; De Bastiani, M.; Peng, W.; Dursun, I.; Sinatra, L.; El-Zohry, A. M.; Hedhili, M. N.; Emwas, A. H.; Mohammed, O. F.; Ning, Z.; Bakr, O. M., Bidentate Ligand-Passivated CsPbI3 Perovskite Nanocrystals for Stable Near-Unity Photoluminescence Quantum Yield and Efficient Red Light-Emitting Diodes. *J. Am. Chem. Soc.* **2018,** *140* (2), 562-565.
6. Chen, P. C.; Liu, M.; Du, J. S.; Meckes, B.; Wang, S.; Lin, H.; Dravid, V. P.; Wolverton, C.; Mirkin, C. A., Interface and Heterostructure Design in Polyelemental Nanoparticles. *Science* **2019,** *363* (6430), 959-964.
7. Green, P. B.; Narayanan, P.; Li, Z.; Sohn, P.; Imperiale, C. J.; Wilson, M. W. B., Controlling Cluster Intermediates Enables the Synthesis of Small PbS Nanocrystals with Narrow Ensemble Line Widths. *Chem. Mater.* **2020,** *32* (9), 4083-4094.
8. Xia, Y.; Xia, X.; Peng, H. C., Shape-Controlled Synthesis of Colloidal Metal Nanocrystals: Thermodynamic versus Kinetic Products. *J. Am. Chem. Soc.* **2015,** *137* (25), 7947-66.
9. Steimle, B. C.; Fenton, J. L.; Schaak, R. E., Rational Construction of a Scalable Heterostructured Nanorod Megalibrary. *Science* **2020,** *367* (6476), 418-424.
10. de Mello Donega, C., Synthesis and Properties of Colloidal Heteronanocrystals. *Chem. Soc. Rev.* **2011,** *40* (3), 1512-46.
11. Liu, L.; Corma, A., Metal Catalysts for Heterogeneous Catalysis: From Single Atoms to Nanoclusters and Nanoparticles. *Chem Rev* **2018,** *118* (10), 4981-5079.
12. Chen, P. Z.; Pollit, L.; Jones, L.; Gu, F. X., Functional Two- and Three-Dimensional Architectures of Immobilized Metal Nanoparticles. *Chem* **2018,** *4* (10), 2301-2328.
13. Kim, T.; Kim, K.-H.; Kim, S.; Choi, S.-M.; Jang, H.; Seo, H.-K.; Lee, H.; Chung, D.-Y.; Jang, E., Efficient and Stable Blue Quantum Dot Light-emitting Diode. *Nature* **2020,** *586* (7829), 385-389.
14. Ozbay, E., Plasmonics: Merging Photonics and Electronics at Nanoscale Dimensions. *Science* **2006,** *311* (5758), 189-93.
15. Howes, P. D.; Chandrawati, R.; Stevens, M. M., Colloidal Nanoparticles as Advanced Biological Sensors. *Science* **2014,** *346* (6205), 1247390.
16. Leshuk, T.; Holmes, A. B.; Ranatunga, D.; Chen, P. Z.; Jiang, Y. S.; Gu, F. X., Magnetic Flocculation for Nanoparticle Separation and Catalyst Recycling. *Environmental Science-Nano* **2018,** *5* (2), 509-519.
17. Zhang, Y.; Malekjahani, A.; Udugama, B. N.; Kadhiresan, P.; Chen, H.; Osborne, M.; Franz, M.; Kucera, M.; Plenderleith, S.; Yip, L.; Bader, G. D.; Tran, V.; Gubbay, J. B.; McGeer, A.; Mubareka, S.; Chan, W. C. W., Surveilling and Tracking COVID-19 Patients Using a Portable Quantum Dot Smartphone Device. *Nano Lett.* **2021,** *21* (12), 5209-5216.
18. Clasky, A. J.; Watchorn, J. D.; Chen, P. Z.; Gu, F. X., From Prevention to Diagnosis and Treatment: Biomedical Applications of Metal Nanoparticle-Hydrogel Composites. *Acta Biomater* **2021,** *122*, 1-25.
19. Markov, I. V., *Crystal Growth for Beginners: Fundamentals of Nucleation, Crystal Growth and Epitaxy*. 3rd edition. ed.; 2017; p xxvii, 604 pages.
20. Ye, X.; Jones, M. R.; Frechette, L. B.; Chen, Q.; Powers, A. S.; Ercius, P.; Dunn, G.; Rotskoff, G. M.; Nguyen, S. C.; Adiga, V. P.; Zettl, A.; Rabani, E.; Geissler, P. L.; Alivisatos, A. P., Single-Particle Mapping of Nonequilibrium Nanocrystal Transformations. *Science* **2016,** *354* (6314), 874-877.

21. Mule, A. S.; Mazzotti, S.; Rossinelli, A. A.; Aellen, M.; Prins, P. T.; van der Bok, J. C.; Solari, S. F.; Glauser, Y. M.; Kumar, P. V.; Riedinger, A.; Norris, D. J., Unraveling the Growth Mechanism of Magic-Sized Semiconductor Nanocrystals. *J. Am. Chem. Soc.* **2021,** *143* (4), 2037-2048.
22. Song, M.; Zhou, G.; Lu, N.; Lee, J.; Nakouzi, E.; Wang, H.; Li, D., Oriented Attachment induces Fivefold Twins by Forming and Decomposing High-energy Grain Boundaries. *Science* **2020,** *367* (6473), 40-45.
23. Lee, J.; Yang, J.; Kwon, S. G.; Hyeon, T., Nonclassical Nucleation and Growth of Inorganic Nanoparticles. *Nature Reviews Materials* **2016,** *1* (8).
24. Takahata, R.; Yamazoe, S.; Koyasu, K.; Imura, K.; Tsukuda, T., Gold Ultrathin Nanorods with Controlled Aspect Ratios and Surface Modifications: Formation Mechanism and Localized Surface Plasmon Resonance. *J. Am. Chem. Soc.* **2018,** *140* (21), 6640-6647.
25. Personick, M. L.; Mirkin, C. A., Making Sense of the Mayhem Behind Shape Control in the Synthesis of Gold Nanoparticles. *J. Am. Chem. Soc.* **2013,** *135* (49), 18238-47.
26. Liz-Marzan, L. M.; Grzelczak, M., Growing Anisotropic Crystals at the Nanoscale. *Science* **2017,** *356* (6343), 1120-1121.
27. Lin, H. X.; Lei, Z. C.; Jiang, Z. Y.; Hou, C. P.; Liu, D. Y.; Xu, M. M.; Tian, Z. Q.; Xie, Z. X., Supersaturation-dependent Surface Structure Evolution: From Ionic, Molecular to Metallic Micro/Nanocrystals. *J. Am. Chem. Soc.* **2013,** *135* (25), 9311-4.
28. Tan, C. S.; Hou, Y.; Saidaminov, M. I.; Proppe, A.; Huang, Y. S.; Zhao, Y.; Wei, M.; Walters, G.; Wang, Z.; Zhao, Y.; Todorovic, P.; Kelley, S. O.; Chen, L. J.; Sargent, E. H., Heterogeneous Supersaturation in Mixed Perovskites. *Adv Sci (Weinh)* **2020,** *7* (7), 1903166.
29. Peng, X.; Wickham, J.; Alivisatos, A. P., Kinetics of II-VI and III-V Colloidal Semiconductor Nanocrystal Growth: "Focusing" of Size Distributions. *Journal of the American Chemical Society* **1998,** *120* (21), 5343-5344.
30. Morin, S. A.; Bierman, M. J.; Tong, J.; Jin, S., Mechanism and Kinetics of Spontaneous Nanotube Growth Driven by Screw Sislocations. *Science* **2010,** *328* (5977), 476-80.
31. Bierman, M. J.; Lau, Y. K.; Kvit, A. V.; Schmitt, A. L.; Jin, S., Dislocation-driven Nanowire Growth and Eshelby Twist. *Science* **2008,** *320* (5879), 1060-3.
32. Zhu, J.; Peng, H.; Marshall, A. F.; Barnett, D. M.; Nix, W. D.; Cui, Y., Formation of Chiral Branched Nanowires by the Eshelby Twist. *Nat Nanotechnol* **2008,** *3* (8), 477-81.
33. Joo, J.; Chow, B. Y.; Prakash, M.; Boyden, E. S.; Jacobson, J. M., Face-selective Electrostatic Control of Hydrothermal Zinc Oxide Nanowire Synthesis. *Nat. Mater.* **2011,** *10* (8), 596-601.
34. Abecassis, B.; Testard, F.; Kong, Q.; Francois, B.; Spalla, O., Influence of Monomer Feeding on a Fast Gold Nanoparticles Synthesis: Time-resolved XANES and SAXS Experiments. *Langmuir* **2010,** *26* (17), 13847-54.
35. Lamer, V. K.; Dinegar, R. H., Theory, Production and Mechanism of Formation of Monodispersed Hydrosols. *Journal of the American Chemical Society* **1950,** *72* (11), 4847-4854.
36. Watzky, M. A.; Finke, R. G., Transition Metal Nanocluster Formation Kinetic and Mechanistic Studies. A New Mechanism when Hydrogen is the Reductant: Slow, Continuous Nucleation and Fast Autocatalytic Surface Growth. *Journal of the American Chemical Society* **1997,** *119* (43), 10382-10400.
37. Thanh, N. T.; Maclean, N.; Mahiddine, S., Mechanisms of Nucleation and Growth of Nanoparticles in Solution. *Chem Rev* **2014,** *114* (15), 7610-30.
38. Hendricks, M. P.; Campos, M. P.; Cleveland, G. T.; Jen-La Plante, I.; Owen, J. S., A Tunable Library of Substituted Thiourea Precursors to Metal Sulfide Nanocrystals. *Science* **2015,** *348* (6240), 1226-30.
39. Hendricks, M. P.; Cossairt, B. M.; Owen, J. S., The Importance of Nanocrystal Precursor Conversion Kinetics: Mechanism of the Reaction between Cadmium Carboxylate and Cadmium Bis(diphenyldithiophosphinate). *ACS Nano* **2012,** *6* (11), 10054-62.
40. Yang, T. H.; Gilroy, K. D.; Xia, Y., Reduction Rate as a Quantitative Knob for Achieving Deterministic Synthesis of Colloidal Metal Nanocrystals. *Chem. Sci.* **2017,** *8* (10), 6730-6749.
41. Sugimoto, T., Preparation of Monodispersed Colloidal Particles. *Advances in Colloid and Interface Science* **1987,** *28* (1), 65-108.
42. Bohren, C. F.; Huffman, D. R., *Absorption and Scattering of Light by Small Particles*. Wiley: New York, 1983; p xiv, 530 p.
43. Maes, J.; Castro, N.; De Nolf, K.; Walravens, W.; Abécassis, B.; Hens, Z., Size and Concentration Determination of Colloidal Nanocrystals by Small-Angle X-ray Scattering. *Chem. Mater.* **2018,** *30* (12), 3952-3962.
44. Haiss, W.; Thanh, N. T.; Aveyard, J.; Fernig, D. G., Determination of Size and Concentration of Gold Nanoparticles from UV-Vis Spectra. *Anal. Chem.* **2007,** *79* (11), 4215-21.
45. Talapin, D. V.; Rogach, A. L.; Haase, M.; Weller, H., Evolution of an Ensemble of Nanoparticles in a Colloidal Solution: Theoretical Study. *J. Phys. Chem. B* **2001,** *105* (49), 12278-12285.
46. Kelly, K. L.; Coronado, E.; Zhao, L. L.; Schatz, G. C., The Optical Properties of Metal Nanoparticles: The Influence of Size, Shape, and Dielectric Environment. *The Journal of Physical Chemistry B* **2003,** *107* (3), 668-677.
47. Yu, W. W.; Qu, L.; Guo, W.; Peng, X., Experimental Determination of the Extinction Coefficient of CdTe, CdSe, and CdS Nanocrystals. *Chem. Mater.* **2003,** *15* (14), 2854-2860.
48. Wang, Y.; Peng, H. C.; Liu, J.; Huang, C. Z.; Xia, Y., Use of Reduction Rate as a Quantitative Knob for Controlling the Twin Structure and Shape of Palladium Nanocrystals. *Nano Lett.* **2015,** *15* (2), 1445-50.
49. Ruditskiy, A.; Zhao, M.; Gilroy, K. D.; Vara, M.; Xia, Y. N., Toward a Quantitative Understanding of the Sulfate-Mediated Synthesis of Pd Decahedral Nanocrystals with High Conversion and Morphology Yields. *Chem. Mater.* **2016,** *28* (23), 8800-8806.
50. Zhou, M.; Wang, H.; Vara, M.; Hood, Z. D.; Luo, M.; Yang, T. H.; Bao, S.; Chi, M.; Xiao, P.; Zhang, Y.; Xia, Y., Quantitative Analysis of the Reduction Kinetics Responsible for the One-Pot Synthesis of Pd-Pt Bimetallic Nanocrystals with Different Structures. *J. Am. Chem. Soc.* **2016,** *138* (37), 12263-70.
51. Bullen, C.; Zijlstra, P.; Bakker, E.; Gu, M.; Raston, C., Chemical Kinetics of Gold Nanorod Growth in Aqueous CTAB Solutions. *Crystal Growth & Design* **2011,** *11* (8), 3375-3380.
52. Jain, P. K.; Lee, K. S.; El-Sayed, I. H.; El-Sayed, M. A., Calculated Absorption and Scattering Properties of Gold Nanoparticles of Different Size, Shape, and Composition: Applications in Biological Imaging and Biomedicine. *J Phys Chem B* **2006,** *110* (14), 7238-48.
53. Park, K.; Drummy, L. F.; Wadams, R. C.; Koerner, H.; Nepal, D.; Fabris, L.; Vaia, R. A., Growth Mechanism of Gold Nanorods. *Chem. Mater.* **2013,** *25* (4), 555-563.
54. Harada, M.; Okamoto, K.; Terazima, M., Diffusion of Gold Ions and Gold Particles During Photoreduction Processes Probed by the Transient Grating Method. *J Colloid Interface Sci* **2009,** *332* (2), 373-81.
55. Park, J. E.; Lee, Y.; Nam, J. M., Precisely Shaped, Uniformly Formed Gold Nanocubes with Ultrahigh Reproducibility in

Single-particle Scattering and Surface-enhanced Raman Scattering. *Nano Lett.* **2018,** *18* (10), 6475-6482.
56. Zheng, Y.; Zhong, X.; Li, Z.; Xia, Y., Successive, Seed-Mediated Growth for the Synthesis of Single-Crystal Gold Nanospheres with Uniform Diameters Controlled in the Range of 5-150 nm. *Particle & Particle Systems Characterization* **2014,** *31* (2), 266-273.
57. Sau, T. K.; Murphy, C. J., Room Temperature, High-Yield Synthesis of Multiple Shapes of Gold Nanoparticles in Aqueous Solution. *J. Am. Chem. Soc.* **2004,** *126* (28), 8648-9.
58. Chiu, C. Y.; Li, Y.; Ruan, L.; Ye, X.; Murray, C. B.; Huang, Y., Platinum Nanocrystals Selectively Shaped Using Facet-specific Peptide Sequences. *Nat. Chem.* **2011,** *3* (5), 393-9.
59. Almora-Barrios, N.; Novell-Leruth, G.; Whiting, P.; Liz-Marzan, L. M.; Lopez, N., Theoretical Description of the Role of Halides, Silver, and Surfactants on the Structure of Gold Nanorods. *Nano Lett.* **2014,** *14* (2), 871-5.
60. Alpay, D.; Peng, L.; Marks, L. D., Are Nanoparticle Corners Round? *The Journal of Physical Chemistry C* **2015,** *119* (36), 21018-21023.
61. Liao, H. G.; Zherebetskyy, D.; Xin, H.; Czarnik, C.; Ercius, P.; Elmlund, H.; Pan, M.; Wang, L. W.; Zheng, H., Facet Development during Platinum Nanocube Growth. *Science* **2014,** *345* (6199), 916-9.

For Table of Contents Only:

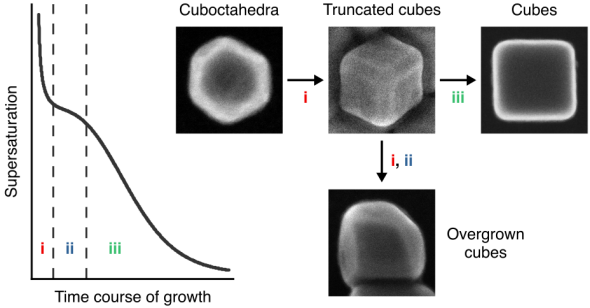

*Supporting Information for*

# Thermodynamic framework elucidating the supersaturation dynamics of nanocrystal growth

Paul Z. Chen[†], Aaron J. Clasky[†], and Frank X. Gu[†,‡,*]

[†]Department of Chemical Engineering & Applied Chemistry, University of Toronto, Toronto, Ontario M5S 3E5, Canada

[‡]Institute of Biomedical Engineering, University of Toronto, Toronto, Ontario M5S 3G9, Canada

Corresponding author: f.gu@utoronto.ca

This supporting information file includes:

S1. Kinetics of NC growth via absorbance and color

S2. Theoretical modeling of NC growth

S3. Development of the simulated supersaturation profile

Supplementary Figures S1–S6

Supplementary Tables S1–S3

## S1. Kinetics of NC growth via absorbance and color

Throughout the entire reaction to grow colloidal Au nanocubes, we took absorbance spectra or color images through the synthesis cuvette (see the Experimental Section for additional information on the Methods). We then analyzed the absorbance or colorimetric kinetics. More specifically, the absorbance kinetics were analyzed using

$$\% \text{ Yield } [Au^0]_{NC,i} = \frac{A_i - \min(\boldsymbol{A})}{\max(\boldsymbol{A}) - \min(\boldsymbol{A})} \tag{S1}$$

where $A_i$ represents the peak absorbance at time $i$ and $\max(\boldsymbol{A})$ and $\min(\boldsymbol{A})$ represent the maximum and minimum peak absorbance, respectively, throughout the reaction. The absorbance spectra were collected using the in-line spectrometer (Methods) and the peak absorbance refers to the highest absorbance value at the peak wavelength between 500-600 nm. The full spectra along the time course of the reaction are shown in Figure 3c.

Since the nanoplasmonic color of the colloidal NCs in this reaction was red (Figure S1a), we used the G value from red-green-blue (RGB) analysis of these color images (ImageJ, version 1.51s; National Institutes of Health, USA). The intrinsic variation among images in white balance across images was normalized using $G_{N,i} = G_{NC,i} - G_{B,i}$, where, for the $i$th image, $G_{N,i}$ represents the normalized G value, $G_{NC,i}$ represents the G value from the colloidal NC formulation, and $G_{B,i}$ represents the G value from the white background. The colorimetric kinetics were then analyzed using

$$\% \text{ Yield } [Au^0]_{NC,i} = \frac{\max(\boldsymbol{G_N}) - G_{N,i}}{\max(\boldsymbol{G_N}) - \min(\boldsymbol{G_N})} \tag{S2}$$

where max $(G_N)$ and min $(G_N)$ represent the maximum and minimum normalized G values for the array of images, respectively, over the reaction. The color images were collected as described in the Methods.

This seed-mediated synthesis had an induction period at the beginning of the reaction. Thus, the absorbance and colorimetric kinetics were fitted to eq (14), in which 100% Yield $[Au^0]_{NC,i}$ was taken to be $C_0 - C_\infty$ if the results were presented in units of concentration rather than % yields. In both cases, equation fitting determined the variable parameters, $k_g$ and $t_0$. Along with the constants summarized in Table S1, these fitted parameters were inputted into eq (15) to develop a continuous, temporal function for eq (8), characterizing the supersaturation dynamics in the reaction.

## S2. Theoretical modeling of NC growth

We used the temporal profile of supersaturation to model the growth of the colloidal Au nanocubes. We inputted the constants listed in Table S1 along with the supersaturation profile derived from the absorbance kinetics into eq (4) to determine the theoretical growth rate at each time $i$. We approximated cuboctahedra, truncated cubes and cubes in our model as pseudospherical NCs, with the radius as the side-to-side length for cube shapes. Since the molar concentration of Br⁻ was low in the growth formulation, $C_\infty$ was estimated based on the equilibrium concentration of $[AuCl_2]^-$, based on a previous approach to determine $C_\infty$ for the monomer of a NC growth reaction.[1] In this seed-mediated growth reaction, $n_{NC}$ was estimated based on the measured peak optical density of the seed ($OD_{peak} = 0.78$ for nanocubes that were diluted 3x in MilliQ water) and the expected particle concentration for the seed, 10-nm nanocuboctahedra[2-3] (https://www.sigmaaldrich.com/CA/en/technical-documents/technical-article/materials-science-

and-engineering/biosensors-and-imaging/gold-nanoparticles), and we took $n_{NC}$ as approximately $2.1 \times 10^{11}$. Since the supersaturation profile and growth rate were discretized in the modeling code, we chose a small time step ($\Delta t$) between each time point in the array to approximate a continuous function for the arrays for the supersaturation profile, growth rate and NC size. The seed size, i.e., the initial size at $t_1 = 0$, was taken to be 10 nm,[2-3] and we multiplied the growth rate at each time $i$ by $\Delta t$ to a determine the NC size at time $i + 1$. This procedure was continued for the duration of the reaction, building a high-resolution growth profile for these NCs. The code generated during this study, including the script used to characterize supersaturation and model growth, is available at GitHub (https://github.com/paulzchen/supersaturation).

## S3. Development of the simulated supersaturation profile

To further investigate how the supersaturation dynamics influenced the growth profile of the NCs, we developed an artificially simulated profile of supersaturation. More specifically, we sought to understand if the complex supersaturation dynamics determined from our framework were appropriate. Thus, we aimed to develop a supersaturation profile that more resembled that expected when considering the LaMer-Dinegar model.[4] Our characterization of the supersaturation dynamics most differed from the LaMer-Dinegar model in the initial phase of rapid monomer consumption (phase **i** in Figure 3e), which arose largely from the dependence ) on $1/r(t)$ for the factor multiplying $R_{NC}(t)$ in eq (8). Hence, we set $r$ as a constant value in this factor to remove the initial phase in the artificially simulated supersaturation profile. The choice of this constant $r$ was made to result in theoretical sizes, as determined by eq (4) and the procedure described above, that matched those of the experimental NCs by the end of growth. This led to the desired artificial supersaturation profile that indeed resembled how supersaturation would be expected to progress according to the LaMer-Dinegar model. The code generated during this study, including the script used to artificially simulate supersaturation and model growth based on this profile, is available at GitHub (https://github.com/paulzchen/supersaturation).

# Supplementary Figures

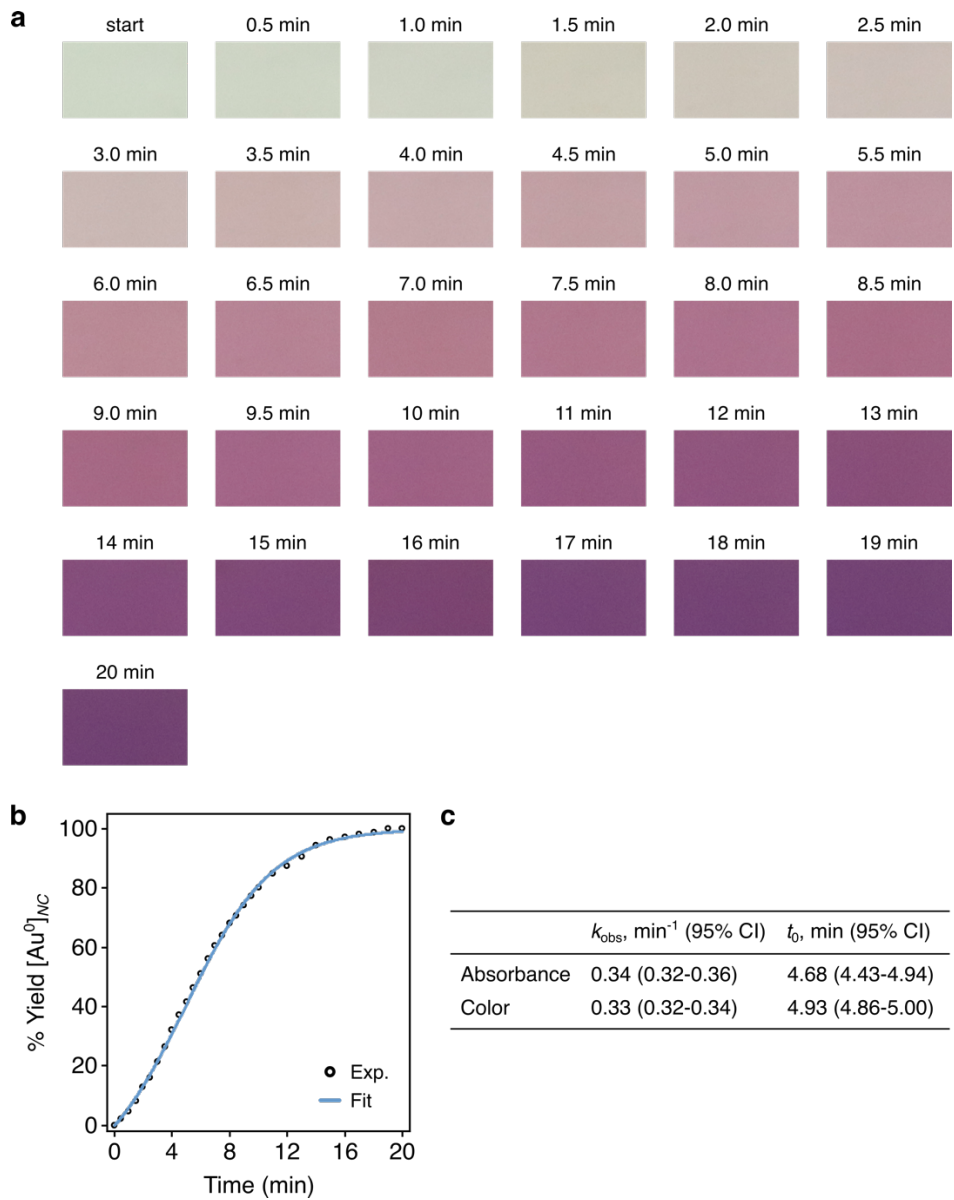

**Figure S1.** Kinetics of NC growth analyzed via nanoplasmonic color development. (a) Time-resolved images of nanoplasmonic color throughout the growth of colloidal Au nanocubes. (b) Kinetics of NC growth analyzed via the colors in (a). The data were fitted to Eq. (11) ($r^2 > 0.99$). (c) Comparison of colloidal NC growth kinetics based on absorbance and colorimetric images. CI, confidence interval.

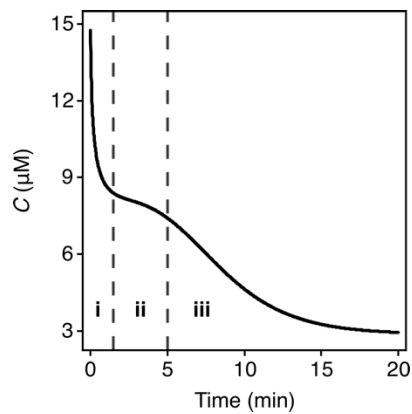

**Figure S2.** Estimated concentrations of monomer throughout growth. This profile was estimated using eq. (1) and the profile of the supersaturation dynamics shown in Figure 3e.

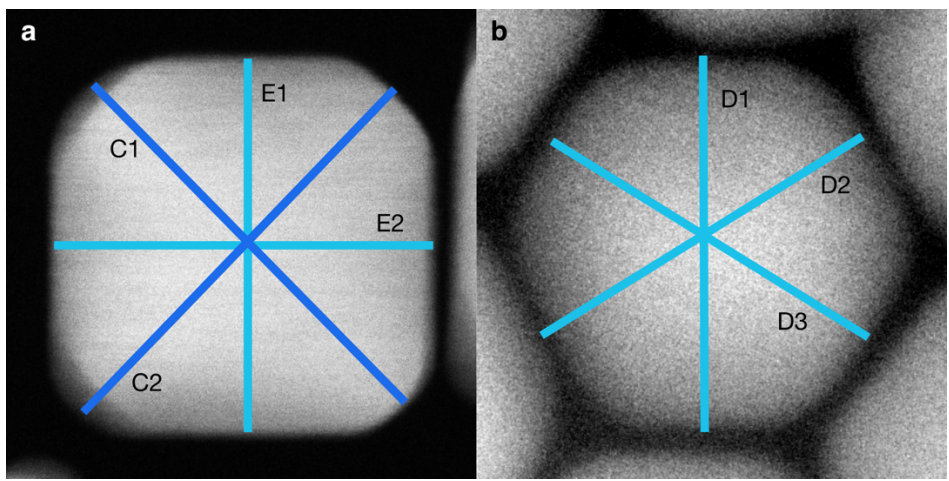

**Figure S3.** Schematic representations of NC sizing measurements from STEM images. (a) Cubes were characterized by corner-to-corner (C) and edge-to-edge (E) distances. (b) Cuboctahedra were characterized by side-to-side (D) distances.

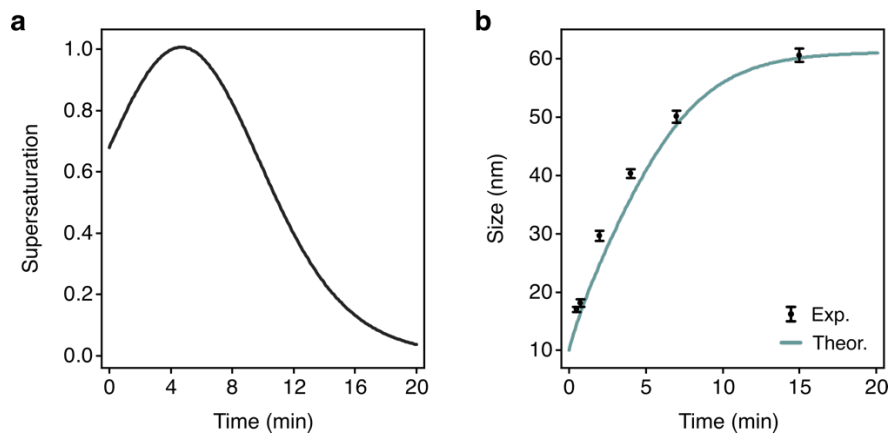

**Figure S4.** Simulated supersaturation without an initial stage of high supersaturation and the resulting predicted growth profile. (a) Profile of simulated supersaturation dynamics. (b) Theoretical growth profile based on the simulated supersaturation profile from (a) and eq (3). The model was overlaid on experimental sizes for the colloidal nanocubes throughout growth. This supersaturation profile was simulated such that the final nanocube size concurred with that which was experimentally observed and theoretically predicted in Figure 3f.

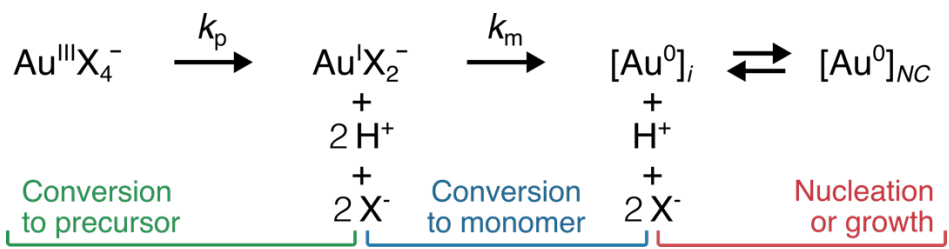

**Figure S5.** Reduction pathway of Au during colloidal NC synthesis. X = halide, which is predominantly $Cl^-$ and small quantities of $Br^-$ in our growth formulation. Both reaction mechanisms for Au reduction, disproportionation and direct reduction, result in the same stoichiometry for the produced monomer, $H^+$ and $X^-$. The addition of $H^+$ and $Cl^-$ shifts the equilibrium, decreasing monomer conversion.

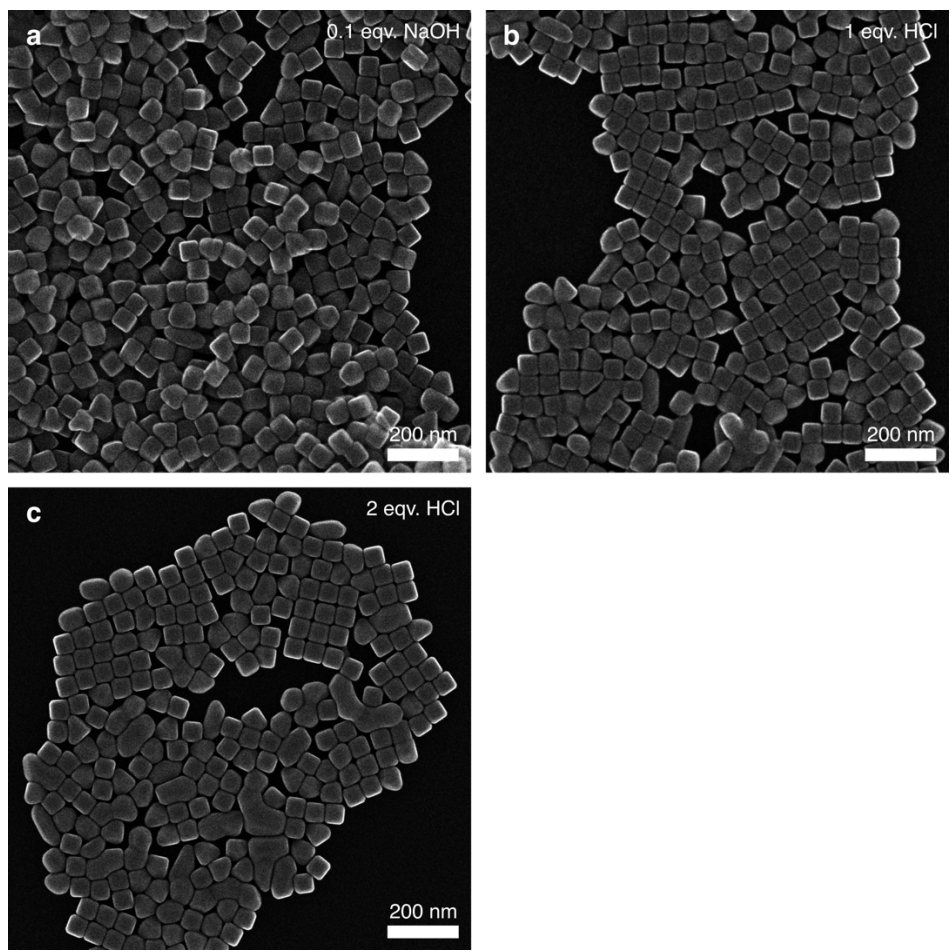

**Figure S6.** Large-area SE-STEM images of Au nanocubes after growth. The micrographs show the NCs grown with (a) 0.1 equivalents (eqv.) NaOH, (b) 1 eqv. HCl or (c) 2 eqv. HCl.

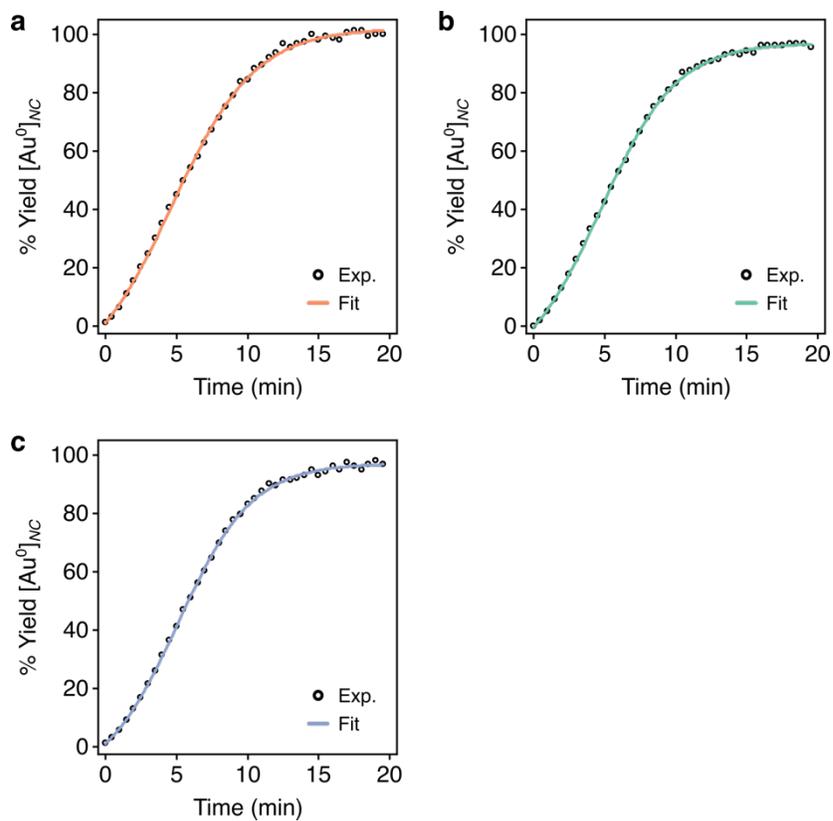

**Figure S7.** Kinetics of NC growth analyzed via the peak nanoplasmonic absorbance. Colloidal Au nanocubes were grown with (a) 0.1 equivalents (eqv.) NaOH, (b) 1 eqv. HCl or (c) 2 eqv. HCl. The data were fitted to Eq. (11) ($r^2 > 0.99$ for each).

## Supplementary Tables

**Table S1.** Parameters used to characterize supersaturation and model colloidal NC growth.

| Symbol | Definition | Units | Value |
|---|---|---|---|
| $a$ | Lattice parameter of Au | Å | 4.085 |
| $v_c$ | Molar volume of Au | cm$^3$ mol$^{-1}$ | 10.3 |
| $C_\infty$ | Saturation concentration of $[Au^0]_i$ | μM | 2.87 ref. [1] |
| $D$ | Bulk diffusion coefficient of monomer/precursor | m$^2$ s$^{-1}$ | $9.0 \times 10^{-10}$ ref. [5] |
| $\nu$ | Adatom vibrational frequency | s$^{-1}$ | $1.0 \times 10^{13}$ ref. [6] |
| $n_{NC}$ | Number of seeds added to growth formulation | - | $2.1 \times 10^{11}$ |
| $d_i$ | Size of nanoseeds added to the growth formulation | nm | 10.0 |
| $T$ | Bulk temperature of growth formulation | K | 293.15 |
| $N_A$ | Avogadro's number | mol$^{-1}$ | $6.022 \times 10^{23}$ |

**Table S2.** Summary of shape evolution throughout colloidal nanocube growth.

| | Shape distribution (%) | | | | | |
|---|---|---|---|---|---|---|
| Time (min) | Cuboctahedra | Truncated cubes | Cubes | Overgrown cubes | Other shapes | Total cubes/ cuboctahedra |
| **0.5** | 96.3 | 0.0 | 0.0 | 0.0 | 3.7 | 96.3 |
| **2.0** | 0.0 | 77.4 | 5.3 | 10.2 | 7.1 | 92.9 |
| **4.0** | 0.0 | 74.0 | 3.3 | 15.0 | 7.7 | 92.3 |
| **7.0** | 0.0 | 34.7 | 35.5 | 21.9 | 7.9 | 92.1 |
| **15** | 0.0 | 8.6 | 60.0 | 24.4 | 7.0 | 93.0 |

**Table S3.** Summary of kinetics and size for colloidal nanocubes when grown with various formulations.

| Sample | $k_{obs}$, min$^{-1}$ (95% CI) | $t_0$, min (95% CI) | Size, nm (SD) |
|---|---|---|---|
| **0.1 eqv. NaOH** | 0.34 (0.32-0.36) | 4.68 (4.43-4.94) | 63.9 (0.89) |
| **1 eqv. HCl** | 0.37 (0.36-0.39) | 4.72 (4.58-4.87) | 62.8 (1.04) |
| **2 eqv. HCl** | 0.38 (0.37-0.40) | 5.10 (4.95-5.25) | 61.2 (1.39) |

# References

1. Morin, S. A.; Bierman, M. J.; Tong, J.; Jin, S., Mechanism and Kinetics of Spontaneous Nanotube Growth Driven by Screw Sislocations. *Science* **2010,** *328* (5977), 476-80.
2. Park, J. E.; Lee, Y.; Nam, J. M., Precisely Shaped, Uniformly Formed Gold Nanocubes with Ultrahigh Reproducibility in Single-particle Scattering and Surface-enhanced Raman Scattering. *Nano Lett.* **2018,** *18* (10), 6475-6482.
3. Zheng, Y.; Zhong, X.; Li, Z.; Xia, Y., Successive, Seed-Mediated Growth for the Synthesis of Single-Crystal Gold Nanospheres with Uniform Diameters Controlled in the Range of 5-150 nm. *Particle & Particle Systems Characterization* **2014,** *31* (2), 266-273.
4. Lamer, V. K.; Dinegar, R. H., Theory, Production and Mechanism of Formation of Monodispersed Hydrosols. *Journal of the American Chemical Society* **1950,** *72* (11), 4847-4854.
5. Harada, M.; Okamoto, K.; Terazima, M., Diffusion of Gold Ions and Gold Particles During Photoreduction Processes Probed by the Transient Grating Method. *J Colloid Interface Sci* **2009,** *332* (2), 373-81.
6. Markov, I. V., *Crystal Growth for Beginners: Fundamentals of Nucleation, Crystal Growth and Epitaxy*. 3rd edition. ed.; 2017; p xxvii, 604 pages.